\documentclass[useAMS,usenatbib]{mnras}

\usepackage{amsmath}
\usepackage{amssymb}
\usepackage{graphicx}
\usepackage{bm}
\usepackage{natbib}
\usepackage{color}

\begin{document}
\title[Origin of Geminga's slow-diffusion halo]{Possible origin of the 
slow-diffusion region around Geminga}

\author[K. Fang et al.]{
Kun Fang$^{1}$\thanks{fangkun@ihep.ac.cn}
Xiao-Jun Bi$^{1,2}$\thanks{bixj@ihep.ac.cn}
Peng-Fei Yin$^{1}$\thanks{yinpf@ihep.ac.cn}
\\
$^{1}$ Key Laboratory of Particle Astrophysics, Institute of High Energy 
Physics, Chinese Academy of Sciences, Beijing 100049, China\\
$^{2}$ School of Physical Sciences, University of Chinese Academy of Sciences, 
Beijing 100049, China\\
}

\maketitle

\begin{abstract}

Geminga pulsar is surrounded by a multi-TeV $\gamma$-ray halo radiated by the
high energy electrons and positrons accelerated by the central pulsar wind 
nebula (PWN). The angular profile of the $\gamma$-ray emission reported by HAWC 
indicates an anomalously slow diffusion for the cosmic-ray electrons and 
positrons in the halo region around Geminga. In the paper we study the possible 
mechanism for the origin of the slow diffusion. At first, we consider the 
self-generated Alfv\'en waves due to the streaming instability of the electrons 
and positrons released by Geminga. However, even considering a very optimistic 
scenario for the wave growth, we find this mechanism DOES NOT work to account 
for the extremely slow diffusion at the present day if taking the proper motion 
of Geminga pulsar into account. The reason is straightforward as the PWN is too 
weak to generate enough high energy electrons and positrons to stimulate strong 
turbulence at the late time. We then propose an assumption that the strong 
turbulence is generated by the shock wave of the parent supernova remnant (SNR) 
of Geminga. Geminga may still be inside the SNR, and we find that the SNR can 
provide enough energy to generate the slow-diffusion circumstance. The TeV 
halos around PSR B0656+14, Vela X, and PSR J1826-1334 may also be explained 
under this assumption.

\end{abstract}


\section{Introduction}
\label{sec:intro}
The well-known $\gamma$-ray pulsar Geminga is surrounded by a multi-TeV 
$\gamma$-ray halo, which was first detected by Milagro 
\citep{2007ApJ...664L..91A}. In late 2017, the High-Altitude Water Cherenkov 
Observatory (HAWC) collaboration further reported the spatially resolved 
observation of the $\gamma$-ray halo \citep{2017Sci...358..911A}. As these 
very-high-energy (VHE) $\gamma$ rays are emitted by electrons and 
positrons\footnote{\textit{Electrons} will denote both electrons and positrons 
hereafter.} mainly through inverse Compton scattering of the cosmic microwave 
background photons, the surface brightness profile of the $\gamma$-ray emission 
can be a good indicator for the propagation of the electrons near the source. 
However, the derived diffusion coefficient of $\sim$60 TeV\footnote{The average 
energy of the $\gamma$ rays observed by HAWC is 20 TeV. Considering both the 
inverse Compton scattering process and the power-law injection spectrum of the 
parent electrons, the average energy of the parent electrons for the 20 TeV 
$\gamma$ rays is $\sim 60$ TeV. We adopt the identical parameters with the 
original paper of HAWC to get this value, including the interstellar radiation 
field and the injection spectral index.} electrons is hundreds times slower 
than the average value in the Galaxy as inferred from the boron-to-carbon ratio 
(B/C) measurements \citep{2016PhRvL.117w1102A}. This is an evidence that the 
diffusion coefficient may be highly inhomogeneous in small scale. Investigating 
the origin of this slow-diffusion region could be meaningful to understand the 
particle propagation near cosmic-ray sources.

A plausible explanation is that the relatively large particle density near the 
source may lead to the resonant growth of Alfv\'{e}n waves, which in turn 
scatter the particles and therefore suppress the diffusion velocity 
\citep{2008AdSpR..42..486P,2013ApJ...768...73M,2016PhRvD..94h3003D}. 
Based on this mechanism, the diffusion coefficient around Geminga can be 
significantly reduced in the case of a hard injection spectrum of electrons and 
a weak ambient magnetic field (this calculation is presented in Appendix 
\ref{sec:app}. See also \citet{2018PhRvD..98f3017E}). However, the precondition 
of this interpretation is that Geminga need to be at rest so that the plenty of 
electrons produced in the early age of Geminga could create a slow-diffusion 
environment. While due to the proper motion of Geminga, it has 
already left 70 pc away from its birthplace \citep{2007Ap&SS.308..225F}, 
which means the observed slow-diffusion region must not be formed in the 
early age of Geminga. The injection power of Geminga in the present day 
should be much weaker than that in the early time, and we will show that the 
diffusion coefficient cannot be remarkably suppressed even in the abscence of 
wave dissipation.

Apart from the self-generated scenario, the slow-diffusion region around 
Geminga could also be a preexisting structure. The diffusion coefficient 
inside a supernova remnant (SNR) should be significantly smaller than that of 
the interstellar medium (ISM), as this region has been swept by the blast wave 
and acquired more turbulent energy. So if Geminga is still inside its 
associated SNR, it could be embedded in a region with small diffusion 
coefficient, which may explain the observed $\gamma$-ray halo. As Geminga may 
have a 70 pc offset from its birthplace at the present day, there are works 
considering that Geminga has already left behind its associated SNR. However, 
we will show below that if the progenitor of Geminga is in a rarefied 
circumstance, the present scale of the SNR could be large enough to include 
Geminga inside. In the light of this explanation, the problems encountered in 
the self-confinement scenario could be avoided. 

In this work, we first test the self-confinement picture in Section 
\ref{sec:self} with very optimistic assumptions, including the disregard of the 
wave dissipation. We consider the impacts brought by the proper motion of 
Geminga, which is an unavoidable factor. Then in Section \ref{sec:snr}, we 
introduce in detail the new interpretation of the inefficient diffusion halo, 
in which the electrons injected by Geminga are diffusing in the turbulent 
environment inside its parent SNR. In Section \ref{sec:dis}, we give some 
further discussion about this topic, including a brief analysis of some other 
TeV inefficient diffusion halos, and an alternative scenario of the preexisting 
kind of origin for the Geminga halo. Finally, we conclude in Section 
\ref{sec:con}.

\section{The self-confined diffusion scenario}
\label{sec:self}
A large density gradient of cosmic-ray particles can induce the streaming 
instability, which may amplify the Alfv\'{e}n waves in background plasma 
\citep{1971ApJ...170..265S}. To derive the diffusion coefficient in the 
vicinity of a source, we must simultaneously solve the equations of particle 
transportation and the evolution of Alfv\'{e}n waves. A full numerical solution 
of the coupled equations is presented in Appendix \ref{sec:app}. Here we only 
show an optimistic scenario for the turbulence growth where the energy loss of 
electrons and the Alfv\'{e}n wave dissipation are ignored. The analysis shows 
clearly why the self-generated mechanism cannot work to stimulate the required 
turbulence.

We neglect the radiative energy loss of electrons, which leads to a larger 
gradient of the number density of electrons. Then the propagation equation is 
expressed as 
\begin{equation}
 \frac{\partial N}{\partial t} - \nabla\cdot(D\nabla N) = Q \,,
 \label{eq:prop}
\end{equation}
where $N$ is the differential number density of electrons, $D$ is the diffusion 
coefficient, and $Q$ is the source function. 

\begin{figure}
 \centering
 \includegraphics[width=0.46\textwidth]{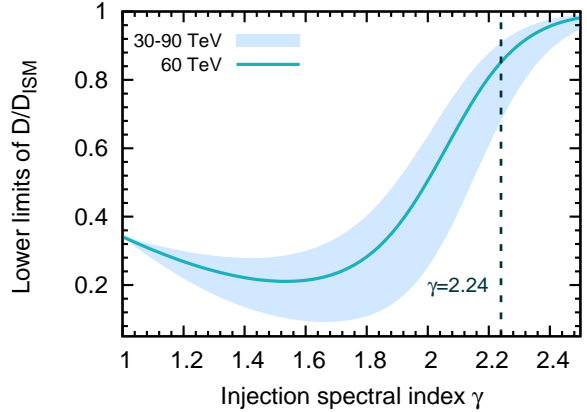}
 \caption{The lower limit of the diffusion coefficient around Geminga under 
the self-confinement scenario. The solid line is the case of 60 TeV, which 
is the mean energy of the parent electrons of the $\gamma$ rays observed by 
HAWC. The band shows the lower limit of diffusion coefficient varying from 30 
to 90 TeV, corresponding to the energy range of HAWC observation. The dotted 
line marks the injection spectral index provided by HAWC.}
 \label{fig:ul}
\end{figure}

The energy density of Alfv\'{e}n waves is denoted with $W$, which is defined by 
$\int W(k)dk=\delta B^2/B_0^2$, where $k$ is the wave number, $B_0$ is the 
mean magnetic field strength, and $\delta B$ is the turbulent magnetic 
field. Here we ignore the wave dissipation and only consider the growth of the 
Alfv\'{e}n waves through streaming instability. The evolution of $W$ can be 
then calculated by 
\begin{equation}
 \frac{\partial W}{\partial t}=\Gamma_{\rm cr}W=-\frac{4\pi 
v_AE_{\rm res}^2}{3B_0^2k}\nabla N(E_{\rm res})\,,
 \label{eq:grow}
\end{equation}
where $\Gamma_{\rm cr}=-4\pi v_AE_{\rm res}^2/(3B_0^2kW)\nabla N(E_{\rm res})$ 
is the growth rate according to \citet{1971ApJ...170..265S}, $v_A$ is the 
Alfv\'{e}n speed, and $E_{\rm res}$ is the energy of electrons satisfying 
$r_g(E_{\rm res})=1/k$, where $r_g$ is the Larmor radius of electrons. 
\citet{2018PhRvD..98f3017E} have proved that the expression of $\Gamma_{\rm 
cr}$ is also applicable for the streaming of electron-positron pairs. The 
diffusion coefficient is related with $W$ by \citep{1971ApJ...170..265S}
\begin{equation}
 D(E_{\rm res})=\frac{1}{3}r_gc\cdot\frac{1}{kW(k)}\,.
 \label{eq:dw}
\end{equation}
Combining Equation (\ref{eq:grow}) and (\ref{eq:dw}), we get 
\begin{equation}
 \frac{1}{D^2}\frac{\partial D}{\partial t}=\frac{4\pi ev_AE}{B_0c}\nabla N\,.
 \label{eq:grow2}
\end{equation}
If Geminga is initially in a typical environment of ISM with $\delta B\ll B_0$, 
the diffusion coefficient along the magnetic field lines should be about 
$(\delta B/B_0)^{-4}$ times larger than the cross-field diffusion coefficient 
\citep{1983RPPh...46..973D}. So the propagation of electrons should be 
initially in a tube of regular magnetic field lines, which corresponds to a 
one-dimensional diffusion. We set $x$ as the coordinate along the regular 
magnetic field lines, and we get the following expression from Equation 
(\ref{eq:prop}) and (\ref{eq:grow2}):
\begin{equation}
 \frac{\partial}{\partial t}\left(N-\frac{B_0c}{4\pi ev_AE}\frac{\partial\ln 
D}{\partial x}\right)=\delta(x)\dot{Q}(t)\,,
 \label{eq:prop2}
\end{equation}
where we assume Geminga is a point-like source, and $\dot{Q}(t)$ is the time 
profile of electron injection. Then for any $x>0$, it can be derived from 
Equation (\ref{eq:prop2}) that
\begin{equation}
 N-\frac{B_0c}{4\pi ev_AE}\frac{\partial\ln D}{\partial x}=0\,.
 \label{eq:prop3}
\end{equation}
Integrating Equation (\ref{eq:prop3}) from $x$ to $\infty$, we finally obtain
\begin{equation}
 D(x)=D_{\rm ISM}\,{\rm 
exp}\left(-\frac{4\pi ev_AE}{B_0c}\int_{x}^{\infty}Ndx'\right)\,,
 \label{eq:limit}
\end{equation}
with $D_{\rm ISM}=D(\infty)$.

Geminga has a proper motion of about 200 km s$^{-1}$ 
\citep{2007Ap&SS.308..225F}, and the direction of motion is suggested to be 
nearly transverse to the line of sight \citep{2003Sci...301.1345C}. These 
indicate that Geminga has left its birthplace for about 70 pc now. Meanwhile, 
the motion of Geminga is almost perpendicular to the Galactic disk 
\citep{1993Natur.361..706G}. This means that Geminga has been cutting the 
magnetic field lines of ISM, as the magnetic field in the Galactic disk is 
dominated by the horizontal component \citep{1994A&A...288..759H}. Thus, the 
electrons injected in the early age of Geminga should have escaped along the 
magnetic field lines which is almost perpendicular to the path of Geminga 
motion and cannot help to generate the present slow-diffusion region.
In other words, the slow-diffusion region around Geminga observed today must be 
formed in the recent age if the region is self-excited by Geminga.

We assume the electrons injected during the last third of the age of Geminga 
(228 kyr$\sim$342 kyr) contribute to the generation of the current 
slow-diffusion region; this should also be an optimistic assumption considering
the very fast energy loss of high energy electrons. The injection time function 
is set to have the same profile with the spin-down luminosity of pulsar, which 
leads to $\dot{Q}(t,E)=\dot{Q}_0(1+t/\tau_0)^{-2}E^{-\gamma}$, where 
$\tau_0=10$ kyr \citep{2017PhRvD..96j3013H}. We assume all the spin-down energy 
of Geminga pulsar is converted to the injected electrons with energy from 1 GeV 
to 500 TeV, to determine the normalization $\dot{Q}_0$. As we have neglected 
the energy loss of electrons, the following relation can be obtained according 
to particle conservation:
\begin{equation}
 2S\int_{x}^{\infty}N(x',E)dx'<\int_{t_1}^{t_2}\dot{Q}(t',E)dt'\,,
 \label{eq:conserve}
\end{equation}
where $t_1=228$ kyr,  $t_2=342$ kyr, and $S$ is the cross-section of the 
magnetic flux tube which is assumed to have a scale of 1 pc. Combining 
Equation (\ref{eq:limit}) and (\ref{eq:conserve}), we can then calculate the 
lower limit of the diffusion coefficient.

\begin{figure}
 \centering
 \includegraphics[width=0.47\textwidth]{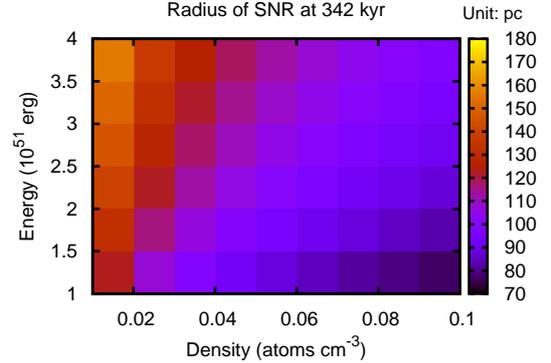}
 \caption{The radius of an SNR at the age of 342 kyr, with varying explosion 
energy and ambient mass density. The calculator provided by 
\citet{2017AJ....153..239L} is applied to get this figure.}
 \label{fig:scale}
\end{figure}

The Alfv\'{e}n speed is decided by $B_0$ and the ion density 
$\rho_i$ as $v_A=B_0/\sqrt{4\pi\rho_i}$. Then Equation (\ref{eq:limit}) 
indicates that $D(x)$ is independent of $B_0$ in our calculation. Considering 
the morphology of the bow-shock structure observed in X-ray 
\citep{2003Sci...301.1345C} and the latest distance measure of Geminga 
\citep{2007Ap&SS.308..225F}, the ISM density around Geminga $\rho_{\rm ISM}$ is 
derived to be 0.02 atoms cm$^{-3}$. Since the ionization around Geminga is very 
high \citep{2003Sci...301.1345C}, we have $\rho_i\approx\rho_{\rm ISM}$. The 
injection spectral index $\gamma$ cannot be well constrained, as HAWC provides 
only the energy-integrated result at present. 

In Figure \ref{fig:ul}, we present the lower limit of diffusion coefficient 
for varying $\gamma$. When $\gamma\approx2.24$ as provided by HAWC, we have 
$D({\rm 60\,TeV})>0.85\,D_{\rm ISM}({\rm 60\,TeV})$. For 60 TeV particles, the 
minimum of the lower limit appears at $\gamma=1.54$, where 
$D({\rm 60\,TeV})>0.21\,D_{\rm ISM}({\rm 60\,TeV})$. However, this is still too 
far from the level of suppression required by HAWC observation, which is only
about $10^{-3}$ of the normal diffusion value in ISM as determined by fitting
the latest B/C value in \citet{2017PhRvD..95h3007Y}. Therefore it is clearly 
shown that the self-confinement mechanism cannot serve as the main reason for 
the suppression of $D(\sim\rm60\,TeV)$ around Geminga.

\begin{figure}
 \centering
 \includegraphics[width=0.45\textwidth]{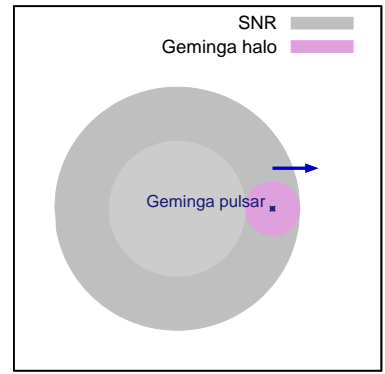}
 \caption{The sketch of the scenario introduced in Section \ref{sec:snr}. The 
arrow denotes the direction of the proper motion of Geminga pulsar. In the 
calculation of Section \ref{sec:snr}, the magnetic field turbulence  is assumed 
to be distributed in the dark-gray shell.}
 \label{fig:sketch}
\end{figure}

\begin{figure*}
 \centering
 \includegraphics[width=0.42\textwidth]{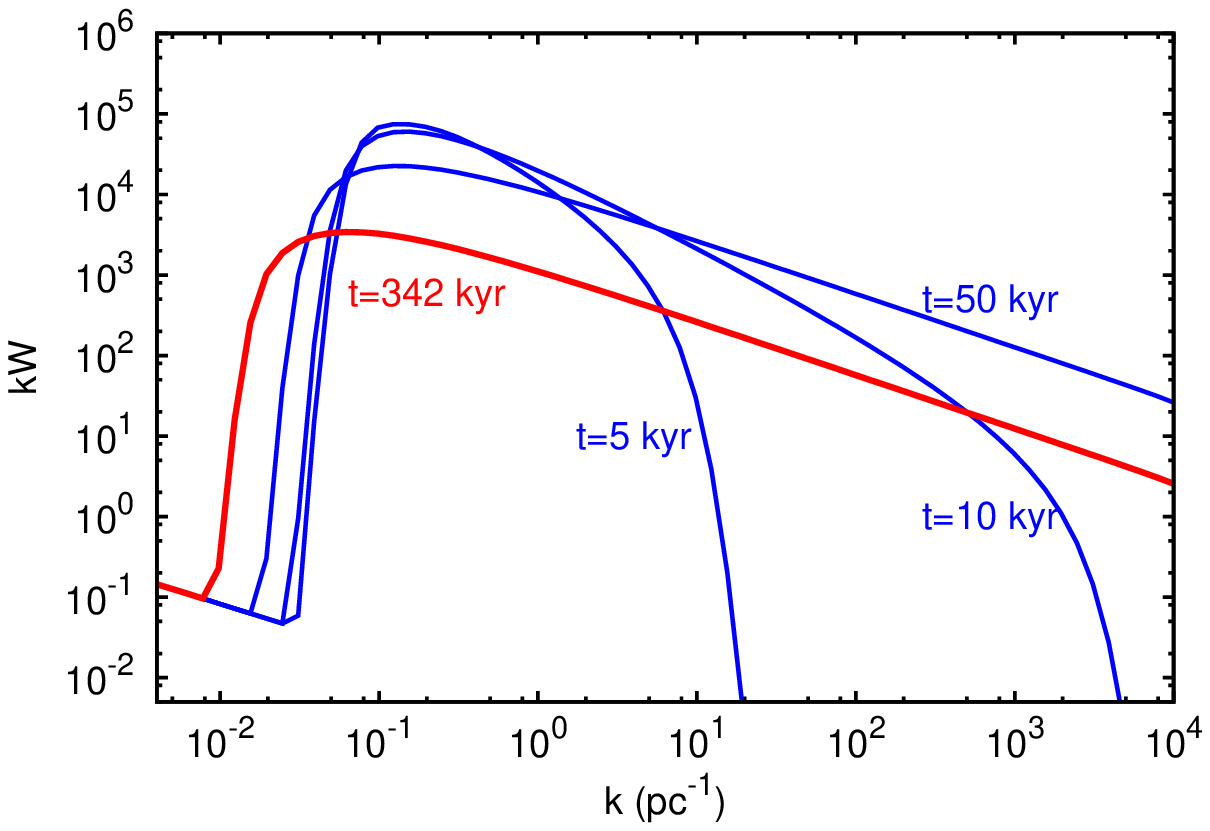}
 \includegraphics[width=0.48\textwidth]{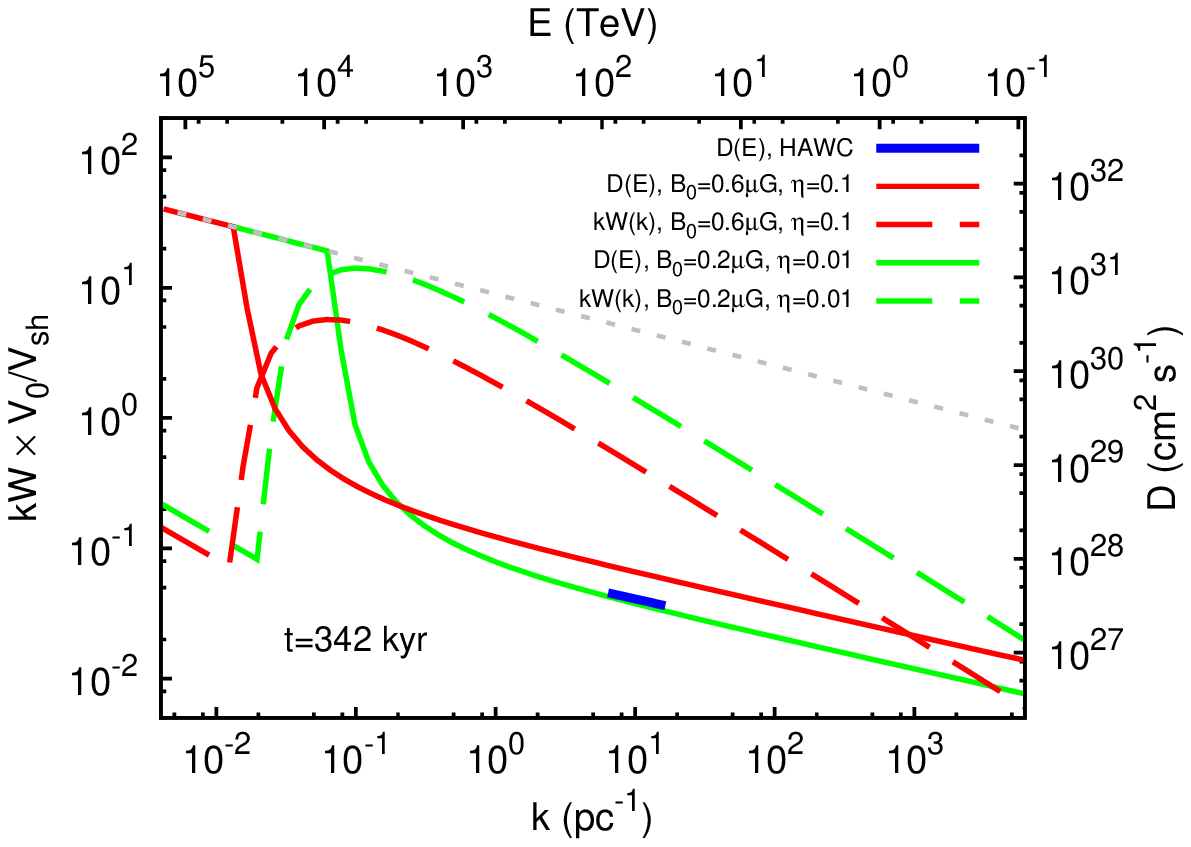}
 \caption{Left: the time evolution of the wave spectrum of the MHD turbulence, 
which is dominated by the wave cascading. Note that in this graph $W$ is not 
diluted with the expansion of the SNR. Right: the present-day turbulence 
spectrum inside the Geminga SNR and the corresponding diffusion coefficient, 
compared with the diffusion coefficient observed by HAWC. The gray dotted line 
is the diffusion coefficient in the ISM. Two different parameter sets are 
adopted: $B_0=0.6$ $\mu$G, $\eta=10\%$ (red);  $B_0=0.2$ $\mu$G, $\eta=1\%$ 
(green).}
 \label{fig:cascade}
\end{figure*}

\section{Electron and positron diffusion inside the SNR}
\label{sec:snr}
In the SNR shock frame, the upstream plasma loses part of kinetic energy 
when streaming through the shock front, and this part of energy is transfered 
into turbulence and thermal energy behind the shock 
\citep{1978MNRAS.182..147B}. Thus, the downstream region can be highly 
turbulent \citep{2007ApJ...663L..41G}, although the turbulence will be 
gradually declined with the long-time evolution of SNRs. So if Geminga 
is still inside its associated SNR, the slow-diffusion region around it may be 
explained. 

We first give an estimate of the possible scale of the Geminga SNR. We 
adopt the calculator provided by \citet{2017AJ....153..239L}. This implement is 
created for modeling the evolution of SNRs, and consistently combines different 
models for different stages of SNR evolution. The SNR dynamic evolution is 
decided by the parameters such as the initial energy of the ejecta $E_0$ and 
the density of the ISM $n_{\rm ISM}$. Figure \ref{fig:scale} shows the radius 
of an SNR at the age of Geminga (342 kyr), with different $E_0$ and $n_{\rm 
ISM}$. The ejecta mass is fixed at 1.4$M_\odot$. For a typical initial energy 
of $1\times10^{51}$ erg, the scale of the SNR can reach $\sim$ 100 pc if the 
ambient density is relatively low with $n_{\rm ISM}<0.05$ cm$^{-3}$. We note 
that Geminga is in the southeast of Monogem Ring on the sky map (in the 
Galactic coordinate), and the distance of Monogem Ring is believed to be 
similar with that of Geminga. The ISM density in the south of Monogem Ring is 
derived to be $0.034$ cm$^{-3}$ \citep{2018MNRAS.477.4414K}, so we assume the 
same ambient density for the parent SNR of Geminga. In this case, the current 
scale of the SNR is 90 pc, if $E_0=10^{51}$ erg.

As mentioned above, Geminga has left its birthplace for about 70 pc now. 
Considering the current radius of the Geminga SNR and the size of the observed 
Geminga halo ($\sim$ 20 pc), we may envisage a scenario in which Geminga has 
been chasing the SNR shock, as presented in Figure \ref{fig:sketch}. Meanwhile, 
the corresponding shock temperature is about $10^5$ K according to the 
calculator of \citet{2017AJ....153..239L}, and the temperature inside the SNR 
should be higher. This is consistent with the high ionization degree around the 
pulsar wind nebula (PWN) of Geminga, as indicated by the measurement of 
the H$\rm \alpha$ luminosity \citep{2003Sci...301.1345C}.

The turbulent energy should be mainly generated in the very early age of an 
SNR, as the shock speed rapidly decreases after the ejecta dominated stage. For 
our parameter setting, the transition age of Geminga SNR from the ejecta 
dominated stage to the Sedov-Taylor stage is $\sim850$ yr, which is negligible 
compared with the current age of Geminga. So it is reasonable to consider a 
burst-like injection for the turbulent energy. For simplicity, we assume that 
the turbulent energy is injected homogeneously into the SNR. Then we write the 
evolution equation of the magnetic field turbulence as
\begin{equation}
\left\{
\begin{aligned}
 & \frac{\partial W}{\partial t}=\frac{\partial}{\partial 
k}\left(D_{kk}\frac{\partial W}{\partial k}\right) \\
 & W(0,k)= Q_W\delta(t)\delta(k-k_0)+W_{\rm ISM}\\
\end{aligned}
\right.\,,
\label{eq:cascade}
\end{equation}
where we assume the evolution is dominated by the turbulent cascading, with the 
Kolmogorov type diffusion coefficient $D_{kk}=0.052\,v_Ak^{7/2}W^{1/2}$ 
\citep{1995ApJ...452..912M}. The wave damping due to the ion-neutral 
interaction is not significant for a high ionization environment 
\citep{1971ApL.....8..189K}. The initial condition is the sum of the injection 
term and the ISM term, and the latter is related to $D_{\rm ISM}$ by Equation 
(\ref{eq:dw}). We assume the injection scale of the MHD turbulence to be 
$l_0=10$ pc, corresponding to $k_0=0.1$ pc$^{-1}$; the size of Geminga can 
indeed reach $\sim$10 pc in the very early age for our case. The normalization 
of the injection term is expressed by 
\begin{equation}
 Q_W=\frac{\eta E_0/V_0}{B^2_0/8\pi}\,,
 \label{eq:qw}
\end{equation}
where $\eta$ is the conversion efficiency of the magnetic field turbulence, and 
$V_0=4\pi l_0^3/3$. \citet{2018MNRAS.479.4526L} argued that the HAWC 
observation of Geminga halo favors an rms magnetic field of 3 $\mu$G. So if the 
outer scale of the Kolmogorov type turbulence is 10 pc,  the mean field $B_0$ 
should be about 0.6 $\mu$G.

We numerically solve Equation (\ref{eq:cascade}) with the finite difference 
method. As Equation (\ref{eq:cascade}) is a non-linear problem, we adopt the 
predictor-corrector method to linearize the difference equation set. 
The wave number $k$ spans a broad range of magnitude in the numerical 
calculation, so we convert $k$ into the logarithmic scale with 
$x=\log_{10}[k\,{\rm (cm^{-1})}]$. The step lengths are set to be $\Delta 
x=0.1$ and $\Delta t=2$ yr to ensure the accuracy of the solution.

The evolution of the wave spectrum is presented in the left of Figure 
\ref{fig:cascade}. The wave spectrum expands in the $k$ space and converges to 
the Kolmogorov type ($W\sim k^{-5/3}$) after the age of $\sim10$ kyr. Then 
the intensity of the spectrum gradually decreases until the present day. 
However, with the expansion of the SNR, the turbulent energy injected in the 
early should be diluted. To compare with the observation, we assume that the 
turbulent energy is now homogeneously distributed in a shell 50$-$90 pc from 
the SNR center, which includes the Geminga halo in. For an old SNR, the mass is 
indeed distributed in the outer part \citep{1988ApJ...334..252C}, and so does 
the turbulence. Then the turbulent energy density is diluted from $W$ to 
$W\times V_0/V_{\rm sh}$, and $V_{\rm sh}$ is the volume of the shell where the 
turbulence is mainly distributed. Note that this is a conservative estimation 
for $W$. In the real case, $W$ is gradually diluted with the expansion of the 
SNR, not instantaneously at the present day. This implies that $D_{kk}$ should 
be smaller compared with the calculation above. As shown in the left of Figure 
\ref{fig:cascade}, the wave spectrum has been decreasing after tens of kyr. A 
smaller $D_{kk}$ will lead to a higher current wave spectrum.

The current diffusion coefficient corresponding to $W\times V_0/V_{\rm 
sh}$ is shown in the right of Figure \ref{fig:cascade}. For the case of 
$B_0=0.6$ $\mu$G, $D(E)$ can be close to the value reported by HAWC with 
$\eta=10\%$, which means the SNR is energetic enough to explain the inefficient 
diffusion environment of Geminga. Recently, \citet{2019ApJ...875..149L} analyze 
the X-ray data of \textit{XMM-Newton} and \textit{Chandra} around the Geminga 
pulsar and derive an upper limit of the diffuse X-ray flux, which corresponds 
to a maximum magnetic field of 1 $\mu$G within $\sim$1 pc around the Geminga 
pulsar. If this conclusion can be extrapolated to a larger neighbourhood of 
Geminga, we may have a maximum $B_0$ of 0.2 $\mu$G. Then in the right of Figure 
\ref{fig:cascade}, we also show the case of $B_0=0.2$ $\mu$G. In this case, the 
theoretical $D(E)$ can accommodate the observed value only with a conversion 
efficiency of 1\% for the magnetic field turbulence. The reason is that $Q_W$ 
can be larger for a smaller $B_0$. Moreover, the cascading time scale is 
approximately $\tau_c\sim k^2_0/D_{kk}\sim k^{-3/2}_0(\eta 
E_{SN})^{-1/2}V_0^{1/2}\rho_i^{1/2}$. The cascading time scale is larger for a 
smaller $\eta$, which results in a stronger current wave spectrum for a fixed 
$Q_W$, as explained in the last paragraph. Thus to explain the observed 
$D(E)$, the required conversion efficiency $\eta$ is positively correlated with 
$B_0$.

\section{Discussion}
\label{sec:dis}
\subsection{Other TeV halos}
The mechanism to explain the slow diffusion around Geminga proposed in the
previous section can be examined in other similar TeV $\gamma$-ray halos
around pulsars. Besides Geminga there are other pulsars that are observed to be 
surrounded by slow diffusion halos in TeV. The spatial profile of $\gamma$-ray 
emission around PSR B0656+14 was reported by HAWC along with that of Geminga, 
and the indicated diffusion coefficient is 5 times larger compared with the 
Geminga case \citep{2017Sci...358..911A}. While unlike Geminga, the associated 
SNR of PSR B0656+14, namely the Monogem Ring, is still observable in X-ray 
\citep{1971ApJ...167L...3B,1996ApJ...463..224P}, as it is much younger than 
Geminga ($\sim$100 kyr). The position of PSR B0656+14 on the sky map is well 
inside the Monogem Ring, while the observation of the PWN of PSR B0656+14 
suggests that the motion of the pulsar is almost parallel to the line of sight 
\citep{2016ApJ...817..129B}. Since Monogem Ring is an extended structure with a 
scale of $\sim80$ pc \citep{2018MNRAS.477.4414K}, the TeV halo of PSR B0656+14 
can still be included by the SNR as long as its radial velocity is not faster 
than 600 km s$^{-1}$. If so, the origin of the slow-diffusion region could also 
be explained by the scenario of Section \ref{sec:snr}. 

Vela X, as the PWN of Vela pulsar, is close to the center of Vela SNR 
\citep{sushch11}. H.E.S.S has detected an extended TeV structure around Vela 
pulsar with a scale of $\sim6$ pc 
\citep{2006A&A...448L..43A,2012A&A...548A..38A}, which is considered to be 
correlated with the X-ray filament \citep{hinton11}. The TeV halo is more 
extended than the X-ray filament, and the derived magnetic field is only 
$\sim4$ $\mu$G, much smaller than that close to the pulsar \citep{hinton11}. So 
it is possible that the TeV structure is produced by the escaping electrons 
that are wandering in the turbulent environment inside the Vela SNR. 
\citet{2018ApJ...866..143H} also indicates that Vela X should be surrounded by 
a slow-diffusion environment, so that its lepton flux at the Earth will not 
conflict with the current experiments. Besides, the anomalously large 
extended TeV halo around PSR J1826-1334 \citep[HESS 
J1825-137,][]{2006A&A...460..365A,2018ApJ...860...59K} could also be ascribed 
to the diffusion of electrons inside its host SNR.

On the other hand, we pay attention to another source PSR B1957+20, 
around which no TeV structure has been detected so far. PSR B1957+20 is an old 
millisecond pulsar, which is definitely traveling in the ISM now. The bow-shock 
PWN associated to the pulsar has been detected by \textit{Chandra} in 
0.3--8 keV \citep{2003Sci...299.1372S,2012ApJ...760...92H}, and the magnetic 
field of PWN is estimated to be 17.7 $\mu$G \citep{2012ApJ...760...92H}. So the 
parent electrons of the X-ray emission should be as high as tens of TeV, which 
means this PWN can indeed accelerate electrons to VHE. Besides, 
\citet{1997MNRAS.291..162A} pointed out that ground based telescopes 
should be able to detect VHE $\gamma$-ray emission around the pulsar if the 
spin-down luminosity of the pulsar is larger than $10^{34}(r/1\,{\rm kpc})^2$ 
erg s$^{-1}$, where $r$ is the distance of the pulsar. For the case of PSR 
B1957+20, $r$ is inferred as 2.5 kpc \citep{2002astro.ph..7156C}, and the 
spin-down luminosity of the pulsar is $7.48\times10^{34}$ erg s$^{-1}$ 
\citep{2005AJ....129.1993M}, meeting the criterion above. All these imply that 
if the accelerated electrons are effectively confined near the source, TeV 
structure ought to be revealed. However, according to the present work, the VHE 
electrons may not be able to bound themselves by the self-generated waves, and 
the turbulence in the ISM is far from adequate to confine the escaping 
electrons, unlike the case inside the SNR. Thus, plenty of electrons might have 
effectively spread out and the non-detection of VHE emission can be understood.

\begin{figure}
 \centering
 \includegraphics[width=0.46\textwidth]{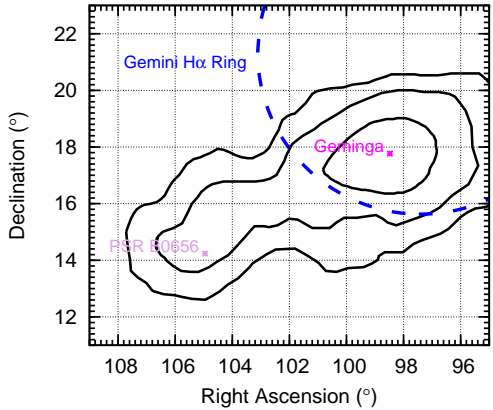}
 \caption{The positions of Geminga and the Gemini H$\alpha$ Ring on the sky 
map. The contours of the HAWC significance map between 1 and 50 TeV ($5\sigma$, 
$7\sigma$, and $10\sigma$) are shown in black \citep{2017Sci...358..911A}. The 
blue dashed line represents the Gemini H$\alpha$ Ring 
\citep{2018MNRAS.477.4414K}.}
 \label{fig:map}
\end{figure}

\subsection{Is Geminga inside a stellar-wind bubble?}
The scenario described in Section \ref{sec:snr} may not be the unique 
possible case for a preexisting slow-diffusion environment. It is also 
possible that Geminga is running into an unrelated turbulent region at present. 
\citet{2007ApJ...665L.139K} discovered a large ring-like structure in H$\alpha$ 
emission that is centered at (97.14$^\circ$, 21.33$^\circ$) in equatorial 
coordinates, dubbed the 'Gemini H$\alpha$ Ring'. As can be seen in Figure 
\ref{fig:map}, the most intense part of the Geminga TeV halo is included by the 
Gemini H$\alpha$ Ring in projection. There is evidence that the Gemini 
H$\alpha$ Ring is interacting with the Monogem Ring 
\citep{2007ApJ...665L.139K}. As the distance of Monogem Ring is estimated to be 
$\sim300$ pc \citep{2018MNRAS.477.4414K}, the Gemini H$\alpha$ Ring should have 
a similar distance. \citet{2018MNRAS.477.4414K} pointed out that the Gemini 
H$\alpha$ Ring is most likely a stellar-wind bubble, as there are several OB 
type stars in the direction of the H$\alpha$ ring with distances of 200-350 pc. 
Meanwhile, the uncertainty of the trigonometric parallax of Geminga is still 
large, and the derived distance ranges from 190 pc to 370 pc 
\citep{2007Ap&SS.308..225F}. Thus, Geminga is possibly inside the stellar-wind 
bubble now, and the shocked wind may provide Geminga a turbulent circumstance. 
Besides, the high ionization and low density environment of Geminga are also 
consistent with the features of a stellar-wind bubble 
\citep{1975ApJ...200L.107C}. 

\section{Conclusion}
\label{sec:con}
We study the possible origin of the slow-diffusion region around Geminga 
observed by HAWC. Considering the proper motion of Geminga, we verify that the 
mechanism of self-generated Alfv\'{e}n waves due to the streaming instability 
cannot work to produce such a low diffusion coefficient even in the most 
optimistic scenario where the energy loss of electrons and the dissipation 
of the Alfv\'{e}n waves are neglected. The reason is simple as Geminga is too 
weak to generate enough high energy electrons at the late age. We get an 
analytical result of the lower limit of the diffusion coefficient, which are at 
most suppressed to about 0.2 times of the ISM value at 60 TeV. This is much 
larger than the value required by the HAWC observation, which derives a 
diffusion coefficient hundreds times smaller than that of the ISM.

We further propose a scenario that the slow diffusion is a preexisting effect, 
as Geminga may still be inside its unidentified parent SNR. We show that if the 
ambient density is low, the scale of the SNR can be large enough to include 
Geminga and the TeV halo inside. We assume that the magnetic field 
turbulence is injected inside the SNR in the very early age of the SNR, and 
then the evolution of the turbulence is dominated by the wave cascading with 
the Kolmogorov form. Our calculation indicates that the diffusion coefficient 
observed by HAWC can be reproduced if 1-10\% of the initial energy of the SNR 
is converted into the magnetic field turbulence, considering the uncertainty of 
the magnetic field strength. Thus, the SNR can provide enough energy to 
explain the slow-diffusion halo around Geminga. We also point out that our 
estimation should be a conservative one, which means the required conversion 
efficiency may be smaller. Another possible interpretation of the slow 
diffusion is also briefly presented, in which Geminga is now running into the 
stellar-wind bubble that creates the Gemini H$\alpha$ Ring.

We further discuss some other sources with TeV halos, such as PSR B0656+14, 
Vela X, and PSR J1826-1334. These cases also favor our new interpretation. 
Recently, \citet{2019arXiv190411536L} propose an alternative scenario that the 
TeV halo of Geminga is not attributed to the strong turbulence, but interpreted 
by the anisotropy diffusion of the electrons along the local regular magnetic 
field which is presumed to be aligned with the line of sight towards Geminga. 
It is still ambiguous if the inefficient diffusion region around pulsar is 
universal or not. As the ground-based Cherenkov instruments have identified 
plenty of VHE $\gamma$-ray sources associated with pulsars (of course many of 
them are PWNe rather than halos produced by escaping electrons)
\citep{2017ApJ...843...40A,2018A&A...612A...1H}, we expect to investigate more 
cases in the future work. Besides, if energy-resolved observation of Geminga 
halo could be provided in the future, it is very helpful to give further 
judgment to the origin of the slow-diffusion region.

\section*{Acknowledgement}
We thank Prof. Hui Li and Dr. Yu-Dong Cui for helpful discussions. We also 
thank the anonymous reviewer for the constructive suggestions. This work is 
supported by the National Key Program for Research and Development 
(No.~2016YFA0400200) and by the National Natural Science 
Foundation of China under Grants No.~U1738209,~11851303.

\appendix

\section{Numerical solution of the self-confinement scenario}
\label{sec:app}
Neglecting the proper motion of Geminga, we numerically solve the complete 
forms of Equation (\ref{eq:prop}) and (\ref{eq:grow}) in the following. The 
radiative cooling of electrons and the damping of Alfv\'{e}n waves are 
considered, then Equation (\ref{eq:prop}) and (\ref{eq:grow}) are rewritten as
\begin{equation}
 \frac{\partial N}{\partial t} - \nabla\cdot(D\nabla N) - 
\frac{\partial}{\partial E}(bN) = Q \,
 \label{eq:prop_b}
\end{equation}
and 
\begin{equation}
 \frac{\partial W}{\partial t}+v_A\nabla W=(\Gamma_{\rm 
cr}-\Gamma_{\rm dis})W\,.
 \label{eq:wave}
\end{equation}
The calculation of the cooling rate $b$ is identical with that in 
\citet{2018ApJ...863...30F}, and the wave cascading is approximated by a 
damping term $\Gamma_{\rm dis}=0.052v_Ak^{3/2}W^{1/2}$ for simplicity 
\citep{2018PhRvD..98f3017E}. In fact, for the ISM that is not infulenced by the 
source, the wave growth and convection can be neglected, while the wave 
dissipation still exists. We universally add a compensatory growth term to keep 
intact the diffusion coefficient far from the source.

We solve Equation (\ref{eq:prop_b}) and (\ref{eq:wave}) iteratively in the 
one-dimensional scenario. For Equation (\ref{eq:prop_b}), we apply the operator 
splitting method to deal with the diffusion operator and the energy-loss 
operator seperately. For each operator, we derive the differencing scheme 
with the finite volume method. This is important especially for the diffusion 
operator, as $D$ can be changed abruptly in space. One may refer 
to \citet{2018ApJ...863...30F} for the details of the differencing schemes. The 
intial value of $N$ is zero everywhere. For the boundary conditions, we set the 
the maximum injection energy to be 500 TeV. The typical scale of the 
Galactic random field is 100 pc, which means the one-dimensional diffusion is 
only valid within this scale around the source. If particles escape farther, 
the diffusion should switch to three-dimensional, and $N$ will sharply 
declines. So we set the spacial outer boundary at 100 pc, namely $N(100\,{\rm 
pc})=0$. The radius of the one-dimensional flux tube is assumed to be 1 pc. As 
to Equation (\ref{eq:wave}), we discretize it with the well-known 
upwind scheme. The initial $W$ is decided by the diffusion coefficient in the 
ISM. To ensure the accuracy of the solutions, $D\Delta t/(\Delta x)^2$ and 
$v_A\Delta t/\Delta x$ should not be much larger than 1, where $\Delta t$ is 
the time step and $\Delta x$ is the radial step for both Equation 
(\ref{eq:prop_b}) and (\ref{eq:wave}). As $D_{\rm ISM}(60\,{\rm 
TeV})\approx3\times10^{30}$ cm$^2$ s$^{-1}$,  we set $\Delta t=0.1$ yr and 
$\Delta x=1$ pc.

\begin{figure}
 \centering
 \includegraphics[width=0.45\textwidth]{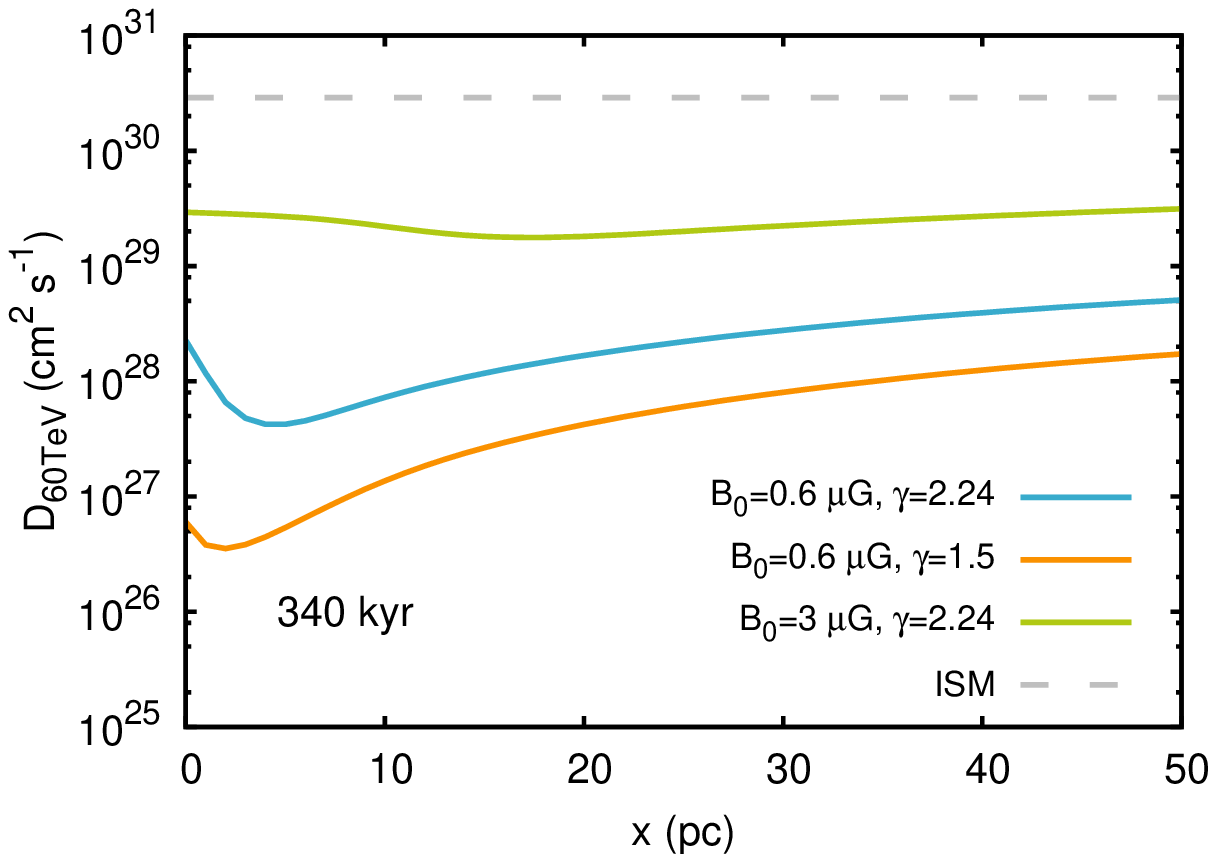}
 \includegraphics[width=0.45\textwidth]{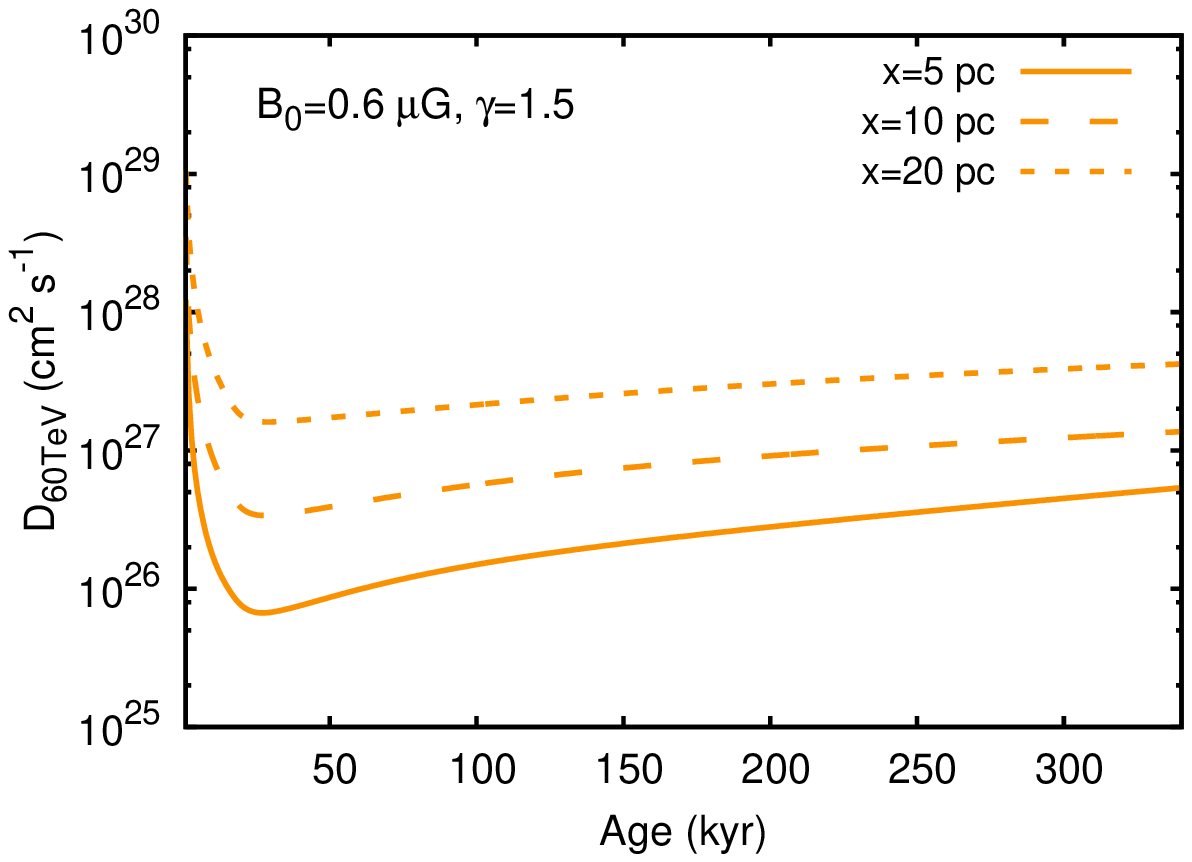}
 \caption{The result of the numerical solution to the self-confined 
propagation of electrons, assuming Geminga is at rest. Top: $D({\rm 60\,TeV})$ 
in the current age of Geminga, for different combinations of $B_0$ and 
$\gamma$. Bottom: the evolution of $D({\rm 60\,TeV})$ at different distances 
from Geminga.}
 \label{fig:evol}
\end{figure}

The spectral index $\gamma$ of the injection spectrum and the mean magnetic 
field $B_0$ are important parameters for the self-confinement scenario. The 
former affects the growth rate of the Alfv\'{e}n waves, while the later is 
decisive for the damping rate of the waves. In the left of Figure 
\ref{fig:evol}, we show the calculated $D({\rm 60\,TeV})$ in the current age of 
Geminga with different $\gamma$ and $B_0$. For the case of harder $\gamma$ 
(1.5) and smaller $B_0$ (0.6 $\mu$G), the diffusion coefficient around Geminga 
is significantly suppressed even in the current age, and the average $D$ within 
20 pc is comparative to the result of HAWC. As can be seen in the right of 
Figure \ref{fig:evol}, the diffusion coefficient declines quickly in the 
early age, then the wave damping dominates and the diffusion coefficient 
gradually rises.

However, in addition to the unrealistic assumption that Geminga is at rest, 
the assumption of one-dimensional diffusion is not always valid. The turbulence 
need to be weak, namely $\delta B\ll B_0$. The right of Figure \ref{fig:evol} 
shows that in the early age of Geminga, the diffusion coefficient can be 
suppressed to very low value, corresponding to strong turbulence. When $\delta 
B$ approaches $B_0$, the diffusion mode should switch to three-dimensional, and 
the wave growth due to streaming instability is significantly reduced compared 
with the case of one-dimensional. This implies that the diffusion coefficient 
cannot be reduced to so low as calculated here, while a self-consistent 
calculation should be complex and beyond the scope of this work.

\end{document}